\documentclass[aps, prb, twocolumn, superscriptaddress, floatfix]{revtex4-2}
\usepackage{etex}

\usepackage{amsmath, amssymb, amsfonts, commath, braket, bm, empheq, array}
\usepackage{mathtools, placeins, standalone}

\usepackage{graphicx, float}
\DeclareGraphicsExtensions{.eps,.jpg,.png,.pdf}
\usepackage{hyperref, url}
\hypersetup{colorlinks=true, linkcolor=red, filecolor=blue, urlcolor=magenta}

\usepackage{dblfloatfix}

\usepackage[utf8]{inputenc}
\usepackage[T1]{fontenc}

\usepackage{xcolor}

\newcommand\as{ \ensuremath{\ket{\mathrm{AS}}} }

\newcommand\esl{ \ensuremath{\ket{\mathrm{ES_1}}} }
\newcommand\esu{ \ensuremath{\ket{\mathrm{ES_2}}} }
\newcommand{\Eth}{\ensuremath{E_{\textrm{Th}}}}	%	Thouless energy
\newcommand{\Ecal}{\ensuremath{\mathcal{E}}}	%	Unfolded energy
\newcommand\gs{ \ensuremath{\ket{\mathrm{GS}}} }

\newcommand\lboro{School of Science, Loughborough University, Loughborough, Leicestershire LE11 3TU, UK}

\newcommand\mean[1]{\ensuremath{\left\langle #1 \right\rangle}}

\newcommand\shn{ \ensuremath{\mathrm{S}} }
\newcommand\tH{\ensuremath{t_\mathrm{H}}}
\begin{document}
%============================================================
\title{Ground state and persistent oscillations in the quantum East model}
\author{Adway Kumar Das}\email{A.K.Das@lboro.ac.uk}
\affiliation{\lboro}
\author{Achilleas Lazarides}\email{A.Lazarides@lboro.ac.uk}
\affiliation{\lboro}

\date{\today}
%============================================================
%	ABSTRACT
\begin{abstract}
	For the 1D quantum East model with open boundary and no global conserved charge, we show that in the limit $s \to -\infty$, the ground state is approximately captured by a simple spin-polarized product state. The polarization angle changes as we approach the quantum phase transition at $s = 0$ where the ground state becomes an exact product state polarized along the X-axis. For $s \to -\infty$, we further identify a low-entanglement excited eigenstate that differs from the ground state only by a $\pi$-rotation of the first spin and is well approximated by a spin-coherent state. For $-1\lesssim s<0$, this spin-coherent state overlaps mainly with two eigenstates separated by a system size-independent energy gap, leading to long-lived coherent oscillations of local observables near the edge. These oscillations originate from boundary physics and are distinct from quantum many-body scars and hypercube-like Fock-space mechanisms.
\end{abstract}
%===========================================================
\maketitle
%===========================================================
\section{Introduction}
%===========================================================
The study of thermalization and its breakdown in isolated quantum systems is a fascinating area of research~\cite{Deutsch:1991ju,Srednicki:1999bo,Rigol:2008bf,Kaufman2016, Wei2018, Dong2025, Borgonovi2016, DAlessio2016, Mori2018, Reimann2008, Linden:2009ii, Das2026}. In the case of ergodic systems, excited eigenstates extend throughout the Hilbert space~\cite{DeLuca2014, Goldstein2010, Neumann2010, DeTomasi2020} and the equilibrium values of local observables are described by the Gibbs ensemble.

One counterexample is many-body localisation (MBL), which occurs in systems with quenched disorder when emergent quasilocal conserved quantities prevent thermalisation~\cite{Basko2006, Pal2010, Nandkishore2015,Serbyn2013, Abanin2019, Chertkov2021,Das2023}. Currently, the stability of the MBL phase in the thermodynamic limit is under intense scrutiny~\cite{DeRoeck2016, Potirniche2019, Suntajs2020, Gopalakrishnan2019, Dumitrescu2019, KieferEmmanouilidis2021, Sels2021, Roy2026}. 

%	MBL without disorder -> East model
A natural question is whether MBL-type physics can occur in translationally invariant systems~\cite{Schiulaz2015, Papic2015, Yao2016,Garrahan_aspects,Schulz2018,Horssen2015,Lan2018,Everest2016}. This question has motivated the study of various kinetically constrained spin-chains without quenched disorder~\cite{Bernien2017, Everest2016, Lan2018, Turner2018, Garrahan_aspects, Laumann2014, Roy2020a, Brighi2024, Royen2024}. Perhaps the prototypical example is the East model where the Hamiltonian can flip a spin of a Fock state only if its neighbor is in a facilitating configuration~\cite{Horssen2015, Garrahan_aspects, Pancotti2020, Rose2022, ValenciaTortora2024, Aditya2025, Menzler2026Arxiv, Menzler2026aArxiv}. This model exhibits a quantum phase transition (QPT) at $s = 0$~\cite{Pancotti2020, Garrahan2007, Banuls2019} where $s$ is the parameter controlling the strength of spin-flips and corresponds to activity in the classical analogue~\cite{Garrahan_aspects}. For $s < 0$, local observables exhibit ballistic spreading and entanglement entropy grows as typically observed in ergodic systems~\cite{Horssen2015}. For $s>0$, local perturbations spread logarithmically with a relaxation time growing exponentially with the system size~\cite{Horssen2015, Menzler2025} and many initial states show slow entanglement growth~\cite{Badbaria2024, Geissler2023}. The kinetic constraint of the East model can be experimentally implemented in Rydberg atoms via a directional antiblockade~\cite{ValenciaTortora2024}.

%================================
\begin{figure}[t!]
	\centering
	\includegraphics[width=\linewidth]{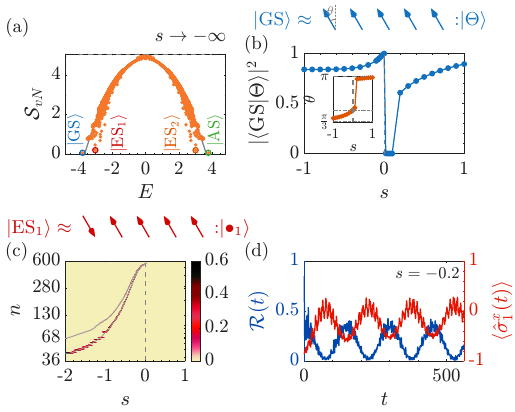}
	\caption{(a)~Bipartite entanglement entropy of energy eigenstates for $L = 16$ and $s\to -\infty$ (Eq.~\eqref{eq:H_East_L}). The dashed line denotes the Page value. % We show 4088 out of $2^{16}$ eigenstates. The solid black curve is quadratic fit. 
		Circles mark the ground state ($\gs$), anti-ground state ($\as$), excited states, $\esl$ and $\esu$ with unusually low entanglement compared to their neighboring states.
		%-----------------------
		(b)~Overlap of the spin-polarized state $\ket{\Theta}$ [$\theta_j = \theta$ in Eq.~\eqref{eq:spin-pol_GS}] with $\gs$ vs.~$s$ for $L = 18$. Inset shows the optimum angle of the individual spins of $\ket{\Theta}$. Above it is a cartoon of $\ket{\Theta}$ approximating $\gs$ for $s \leq 0$. 
		%-----------------------
		(c)~Overlap of $\ket{\bullet_1}$ with eigenstates ordered in energy ($n$ is the index) for $L = 14$ [colorbar denotes $|\mean{\Psi_n|\bullet_1}|^2$]. Note how $\ket{\bullet_1}$ for negative-enough $s$ mainly overlaps a single eigenstate, consistent with its large overlap with $\esl$ at $s\to-\infty$, while it overlaps strongly with two eigenstates for $-1\lesssim s<0$ [gray curves are guide to the eye],
		leading to the oscillatory dynamics in panel (d). Above is a cartoon of the spin-coherent state, $\ket{\bullet_1}$ which approximates the eigenstate $\esl$ and differs from $\gs$ [see panel (b)] only at the first spin. 
		%-----------------------
		(d)~Time evolution of the return probability $\mathcal{R}(t)$ (blue curve) and the magnetization along the X-direction at the first site, $\langle \hat{\sigma}_1^x(t) \rangle$ (red curve), for $s = -0.2$ and $L = 16$, where the initial state is $\ket{\bullet_1}$.
	}
	\label{fig1}
\end{figure}
%================================

Here, we focus on the 1D East model with open boundary condition which displays a QPT at $s=0$~\cite{Pancotti2020, Menzler2025}.
We first study the limit $s\to-\infty$, where we find an accurate analytical approximation to the ground state $\gs$ and to a special higher-energy, low-entropy eigenstate which we denote $\esl$ [Fig.~\ref{fig1}(a)]%lying above it
--this state differs from the $\gs$ only by a single spin rotation at the first site of the chain. Both of these approximations are spin-coherent states [Eq.~\eqref{eq:def_spin-cohere}], locally correlated and with vanishing entanglement. %, with the $\esl$ being an edge mode in that it differs from the $\gs$ only at the very edge. 
We then study the region $-\infty<s<0$, where we find that $\gs$ retains the spin-polarization with $s$-dependent orientation
% remains approximately of the same form
[see Fig.~\ref{fig1}(b) for the overlap of our ansatz with $\gs$] while the spin-coherent approximation of $\esl$ eventually (for $-1\lesssim s\lesssim 0$) develops overlaps with two eigenstates with a system size-independent energy gap between them [Fig.~\ref{fig1}(c)]. This implies that the low-entanglement state displays long-lived coherent oscillations as shown in Fig.~\ref{fig1}(d).

%---------------------------------
%   Organization
The organization of this paper is as follows: in Sec.~\ref{sec_model}, we introduce the East model and look at the low-energy physics in the limit $s\to-\infty$ in Sec.~\ref{sec_s_to_minus_infty}. We extend our study to finite $s<0$ in Sec.~\ref{sec_finite_s} and show that the spin-coherent approximation of $\esl$ undergoes long-lived oscillations. Our concluding remarks are given in Sec.~\ref{sec_conclusion}.

%===========================================================
%	Model
\section{Model}\label{sec_model}
%===========================================================
The East model with open boundary condition has two distinct symmetries leading to $2L + 1$ decoupled subspaces in the Hilbert space~\cite{Pancotti2020}. In any of those subspaces apart from the $1\times 1$ subspace spanned by $\ket{0}^{\otimes L}$, the East model has the following irreducible Hamiltonian
\begin{align}
	\begin{split}
		\hat{H}_\mathrm{East} &= \frac{\hat{\mathbb{I}} + \hat{N}}{2} - \frac{e^{-s}}{2} \del{ \hat{\sigma_1}^x + \sum_{j = 1}^{L-1} \hat{n}_j\hat{\sigma}_{j+1}^x + \hat{n}_L } 
	\end{split}
	\label{eq:H_East_EBC}
\end{align}
where $\hat{\sigma}_j^{x, y, z}$ are the Pauli operators and $\hat{N} \equiv \sum_{j = 1}^L \hat{n}_j$ with $\hat{n} = \frac{\hat{\mathbb{I}} - \hat{\sigma}^z}{2}$ being the number density operator having eigenstates $\ket{0} = \begin{pmatrix}
	1 \\ 0
\end{pmatrix}$ and $\ket{1} = \begin{pmatrix}
	0 \\ 1
\end{pmatrix}$. The $\frac{\hat{N}}{2}$ term is a chemical potential $\mu\sum_j n_j$ where we set $\mu=\frac{1}{2}$~\cite{Horssen2015}. The single-spin terms inside the brackets correspond to the fact that the chain of $L$ qubits governed by $\hat{H}_\mathrm{East}$ is coupled to two fictitious qubits: $\hat{\sigma}_1^x$ arises from the $0$th qubit in state $\ket{1}$ and $\hat{n}_L$ from the $(L+1)$the qubit in the state $\ket{+}$.

%===========================================================
\subsection{$s\to-\infty$}\label{sec_s_to_minus_infty}
%===========================================================
We begin our study in the limit $s\to -\infty$, where the Hamiltonian in Eq.~\eqref{eq:H_East_EBC} reduces to
\begin{align}
	\hat{H}_{-\infty} = -\hat{\sigma}_1^x - \sum_{j = 1}^{L-1} \hat{n}_j\hat{\sigma}_{j+1}^x - \hat{n}_L.
	\label{eq:H_East_L}
\end{align}
The density of states of the above Hamiltonian is close to a Gaussian distribution (Appendix~\ref{app:DOS}). Figure~\ref{fig1}(a) shows the bipartite von Neumann entanglement entropy~\cite{Haque2022, DeTomasi2020, Abanin2019} of the eigenstates of $\hat{H}_{-\infty}$ as a function of energy for $L = 16$, which has a rainbow shape typical of interacting many-body systems, with the exception of a pair of peculiar excited states (marked by circles) with entanglement much lower than that of their neighboring eigenstates. These special excited states, denoted by $\esl$ and $\esu$, are closely related to the ground state as we will show later.

%----------------------------------
%	Fractal dimensions of |GS>
To characterize the structure of the ground state $\gs$ in the Fock space, we compute the fractal dimensions $D_q$ from the scaling of the $q$th moments of its wavefunction intensities~\cite{Soukoulis1984,Rodriguez2010,Evers2008,Das2025b}. For ergodic states $D_q=1$ for all $q$ while localized states have $D_q=0$; generally, $D_q$ decreases monotonically with $q$ and so is bounded below by $D_\infty$~\cite{janssen_mutifractal_1994,Lindinger2019,Lakshminarayan2008}. In the $Z$-basis we find $0 < D_q^z < 1$, decreasing with $q$, with finite $D_\infty^z \approx 0.47$~[Appendix~\ref{app:fractal-dims-limit}], indicating that $\gs$ is multifractal: it occupies an extensive yet vanishing fraction of Hilbert space, with distinct scaling exponents for different moments~\cite{Kravtsov2015,DeLuca2014,Das2022,Roy2023}. Multifractality is also present in the $X$-basis, but with $D_q^z > D_q^x$ for all $q>0$, implying greater delocalization in the $Z$-basis. This contrasts with the $s\to\infty$ limit, where the ground state $\ket{0}^{\otimes L}$ is localized in $Z$ and fully extended in $X$ basis ($D_q^x=1$ for all $q>0$).

In real space, local observables such as $\mean{\hat n_j}$ and $\mean{\hat\sigma_j^x}$ are approximately uniform across the chain~[Fig.~\ref{fig:East_L_ground}(a)], despite the multifractality in Fock space. The bipartite entanglement entropy remains small and essentially independent of system size, $\mathcal S_{vN}\approx 0.0472$, consistent with an area law as expected for the ground state of a local Hamiltonian, but here coexisting with nontrivial multifractality in the configuration space~\footnote{We note that our numerical results cannot exclude a logarithmic dependence of $\mathcal S_{vN}$ on $L$, but this does not affect our conclusions.}.

%----------------------------------
%	|GS> ansatz
Overall, then, $\gs$ has low entanglement entropy and is nonergodic in Fock space but extended in real space. This leads us to conjecture that the $\gs$ is approximated by a spin-coherent state:
\begin{align}
	\ket{\cbr{\theta_j}} = \otimes_{j = 1}^{L} \ket{\theta_j},\quad \ket{\theta_j} = \cos\frac{\theta_j}{2}\ket{1} + \sin\frac{\theta_j}{2}\ket{0}.
	\label{eq:def_spin-cohere}
\end{align}
The spin-polarized state ($\theta_j = \theta$ for all values of $j$) is denoted by $\ket{\Theta}$. The energy of the spin-coherent state w.r.t.~the Hamiltonian in Eq.~\eqref{eq:H_East_L} is
\begin{equation*}
	E_{\{\theta_j\}} = -\sin\theta_1 - \sum_{j = 1}^{L-1} \frac{1+\cos\theta_j}{2}\sin\theta_{j+1} - \frac{1+\cos\theta_L}{2}.
\end{equation*}
This is minimized by $\theta_j\approx \frac{\pi}{3}$ for all $j$ and $L$; this is straightforward to see in the case of periodic boundary conditions, where choosing $\theta_j=\theta$ gives an energy $-\frac{L}{2} \sin\theta (1 + \cos\theta)$, minimized at $\theta=\frac{\pi}{3}$. Thus, the ground state is well-approximated by the ansatz
\begin{align}
	\gs_{s\to-\infty} \approx \ket{\frac{\pi}{3}}^{\otimes L}.% \del{\frac{\sqrt{3}}{2}\ket{1} + \frac{1}{2}\ket{0} }^{\otimes L}.
	\label{eq:spin-pol_GS}
\end{align}
The energy of this state is a good approximation to that of the exact $\gs$~[Fig.~\ref{fig:East_L_ground}(c)] for all accessible system sizes (energy per site of $\ket{\Theta}$, $-3\sqrt{3}/8\approx -0.650$, lies above that of the exact ground state $\approx -0.669$). Additionally, our ansatz reproduces all the qualitative features of the ground state: $\bra{\frac{\pi}{3}}\hat{n}\ket{\frac{\pi}{3}} = \frac{3}{4}$ and $\bra{\frac{\pi}{3}}\hat{\sigma}^x\ket{\frac{\pi}{3}} = \frac{\sqrt{3}}{2}$, both of which are close to the expectation values w.r.t.~the ground state~\cite{Banuls2019}. We analytically obtain the fractal dimensions of $\ket{\Theta}$ for $\theta = \frac{\pi}{3}$~[Appendix~\ref{app:fractal-dims-limit}]
\begin{align}
	\begin{split}
		D_q^z &= \frac{\log_2(1+3^q) - 2q}{1 - q}\\
		D_q^x &= \frac{\log_2(\sqrt{3}+1)^{2q} (1 + (\sqrt{3} - 2)^{2q}) - 3q}{1-q}% \frac{2\del{\log_2(\sqrt{3}+1) - 3/2}q + \log_2(1 + (\sqrt{3} - 2)^{2q})}{1-q}
	\end{split}
	\label{eq:Dq_spin_cohere}
\end{align}
which are consistent with the ground state being more delocalized over the $Z$-basis compared to $X$-basis. They are the same for the excited state $\esl$ and also in good agreement with exact results for that [see Fig.~\ref{fig3}(c)].

The overlap of the ansatz from Eq.~\eqref{eq:spin-pol_GS} with the $\gs$ is large for all accessible sizes, $|\mean{\mathrm{GS}| \Theta}|^2 = 0.918,\ 0.888,\ 0.859,\ 0.845$ for $L = 10, 14, 18, 20$, but decreases exponentially with system size $L$. This is because the ground state has low, $L$-independent entanglement entropy and the spin-coherent state approximately captures $\gs$ per site with fidelity slightly less than unity. The overlap of the product state $\ket{\Theta}$ with $\gs$ is bounded by the $L$th power of the per site fidelity, hence, must vanish in the thermodynamic limit ($L \to \infty$).

%----------------------------------

Dynamical support for the strong overlap of the spin-polarized ansatz with the ground state is furnished by studying its return probability
\begin{align}
	\mathcal{R}(t)\equiv \abs{\langle \Theta|\Theta(t) \rangle}^2,\qquad \ket{\Theta(t)}=e^{-iHt}\ket{\Theta}.
\end{align}

In ergodic systems, $\mathcal{R}(t)$ typically relaxes to a plateau after the Heisenberg time $\tH\sim \delta^{-1}$~\cite{Schiulaz2019,VallejoFabila2024,Das2022,Das2022a,Das2023a,VallejoFabila2025,Roy2025} with $\delta$ being the mean level spacing. Here, however, $\mathcal{R}(t)$ remains $\mathcal{O}(1)$ even for $t \gg \tH$ (with $\tH\sim \mathcal{O}(2^L/\sqrt{L})$), reflecting that $\ket{\Theta}$ has a large overlap with the true ground state for the accessible sizes~[Fig.~\ref{fig:East_L_ground}(d)]. This is in agreement with our previous picture of the $s\to-\infty$ ground state as a tilted collective spin for finite system sizes.

In conclusion, the ground state in the $s\to-\infty$ limit is well-approximated by a spin-polarized state, which is the first important result of this work. We now use this result to obtain the anti-ground state. 

%----------------------------------
%	|AS> ansatz
The Hamiltonian $\hat{H} = \hat{H}_{-\infty} + \hat{n}_L$ has terms with an odd number of $\sigma^x$ operators and anti-commutes with the $Z$ parity operator $\hat{\mathcal P}=\bigotimes_j \hat{\sigma}_j^z$, $\{\hat{H}, \hat{\mathcal P}\}=0$~\cite{Bhore2023, Nicolau2026}. This chiral symmetry causes spectral pairing~\footnote{It also causes the Fock-space connectivity to be bipartite, so that even and odd parity sectors couple only to each other.}: if $\hat H\ket{\Psi_n}=E_n\ket{\Psi_n}$, then $\hat{\mathcal P}\ket{\Psi_n}$ is also an eigenstate with energy $-E_n$.

Due to the boundary density operator, $\hat{n}_L$, $\mathrm{Tr} \hat{H}_{-\infty} \neq 0$ and $\{\hat{H}_{-\infty}, \hat{\mathcal P}\} = \bigotimes_{j = 1}^{L-1} \hat{\sigma}_j^z \otimes (\mathbb{I}_L - \hat{\sigma}_L^z)$. Nevertheless, a constant shift to the Hamiltonian shows the following chiral symmetry
\begin{align}
	\cbr{ \hat{H}_{-\infty} + \frac{\mathbb{I}}{2}, \bigotimes_{j=1}^{L-1} \hat{\sigma}_j^z \hat{\sigma}_L^y } = 0.
\end{align}
As the spectrum of $\hat{H}_{-\infty} + \frac{\mathbb{I}}{2}$ is symmetric about zero energy, the energy spectrum of $\hat{H}_{-\infty}$ is symmetric about $E = -\frac{1}{2}$, i.e.~for each energy $E_n$ of $\hat{H}_{-\infty}$, there is a corresponding energy $-E_n-1$. Thus, applying the chiral symmetry operator $\bigotimes_{j=1}^{L-1} \hat{\sigma}_j^z \hat{\sigma}_L^y$ on the ground-state of $\hat{H}_{-\infty}$, we get the corresponding anti-ground state. Note that $\hat{\sigma}_j^z\ket{\frac{\pi}{3}} = \ket{\frac{5\pi}{3}}$ and $\hat{\sigma}_j^y\ket{\frac{\pi}{3}} = i\ket{\frac{10\pi}{3}} = -i\ket{\frac{4\pi}{3}}$. As $\hat{H}_{-\infty}$ has a real symmetric matrix representation in the Z-basis, we ignore the overall phase factor and posit the following ansatz for the anti-ground state
\begin{align}
	\as_{s\to-\infty} \approx \ket{\frac{5\pi}{3}}^{\otimes L-1} \ket{\bullet}, \qquad \ket{\bullet}\equiv\ket{\frac{4\pi}{3}}.
	\label{Eq:AS_ansatz}
\end{align}
Compared to the ansatz for the $\gs$, this state has the opposite $X$-magnetisation and the same fractal dimensions in the Fock space. Since it is the image of $\ket{\Theta}$ under the chiral symmetry, which maps $\gs$ to $\as$, its overlap with the true anti-ground state is exactly that of $\ket{\Theta}$ with $\gs$ for every $L$, $|\braket{\mathrm{AS}|\frac{5\pi}{3}}^{\otimes L-1} \ket{\bullet} |^2 = |\mean{\mathrm{GS}|\Theta}|^2$~\footnote{The ansatz for anti-ground state is orthogonal to that of the ground state, as the two eigenstates must be: $\ket{\frac{5\pi}{3}}$ has overlap $-\frac{1}{2}$ with $\ket{\frac{\pi}{3}}$, giving an exponentially small bulk contribution, and $\ket{\bullet}$ is orthogonal to $\ket{\frac{\pi}{3}}$.}, see Fig.~\ref{fig:East_L_anti-ground}.
%===========================================================
%	Finite s < 0
\subsection{Finite $s < 0$}\label{sec_finite_s}
%===========================================================

%================================
\begin{figure}[t]
	\centering
	\includegraphics[width=\linewidth]{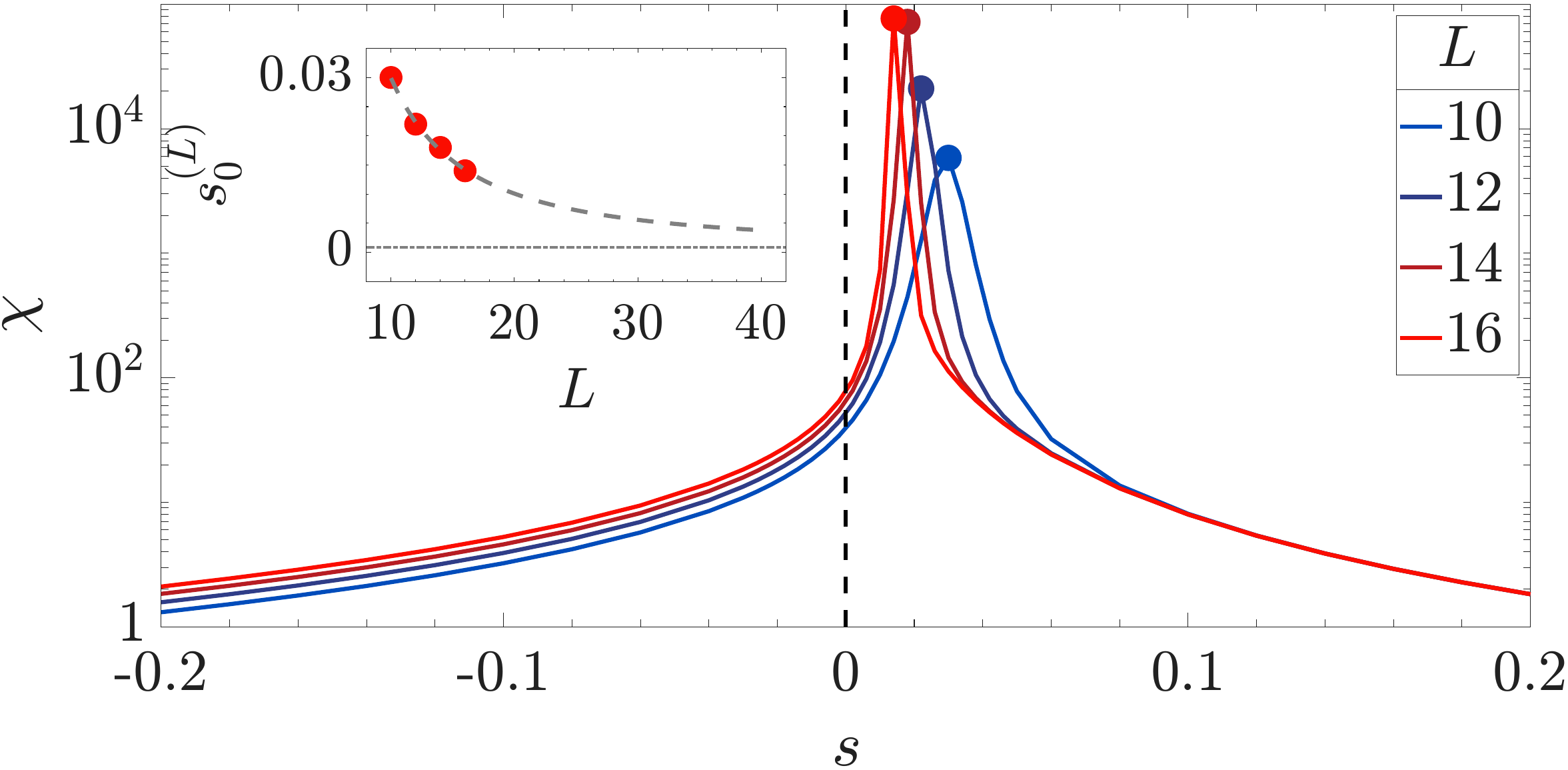}
	\caption{Fidelity-susceptibility $\chi$ of the ground state w.r.t.~$s$ for system sizes $L = 10$--$16$. Inset shows the position $s_0^{(L)}$ of the maximum of $\chi$ vs.~$L$; the dashed line is a power-law fit, $s_0^{(L)} = s_0 + a L^{-\nu}$ with $\nu = 1.66$, which extrapolates to $s_0 = 0.001$ (dash-dotted line).
	}
	\label{fig2}
\end{figure}
%================================

We now extend our study of the ground state to finite $s>-\infty$ [Eq.~\eqref{eq:H_East_EBC}]. As the East Hamiltonian can be expressed as $\hat{H}_{-\infty} + e^{s} (\hat{\mathbb{I}} + \hat{N})$ up to an overall prefactor, the QPT is reflected in $\mathcal{F}(s)$, the fidelity of the ground state w.r.t.~an infinitesimal change in the control parameter, $s\to s+\delta s$~\cite{Sachdev1999Book}
\begin{align}
	\mathcal{F}(s) = \abs{\mean{\mathrm{GS}(s)|\mathrm{GS}(s+\delta s)}} = 1 - \frac{(\delta s)^2}{2}\chi(s) + \mathcal{O}(\delta s^3),
	\label{eq:fidelity}
\end{align}
where $\chi$ is the fidelity-susceptibility of the ground state~\cite{Zanardi2007, You2007, Gu2010, Swietek2025}. In Fig.~\ref{fig2}, we show $\chi$ as a function of $s$ for different system sizes. For a system size $L$, the fidelity-susceptibility has a maximum at $s = s_0^{(L)}$ which is plotted in the inset of Fig.~\ref{fig2} as a function of $L$. For all accessible system sizes the maximum lies at small positive $s$, 
% $s_0^{(L)} = 0.030,\ 0.022,\ 0.018,\ 0.014$ for $L = 10,\ 12,\ 14,\ 16$, 
and approaches $s=0$ as $L$ grows; a power-law fit, $s_0^{(L)} = s_0 + a L^{-\nu}$, gives $\nu = 1.66$ and extrapolates to $s_0 = 0.001$ consistent with the QPT at $s=0$ (the Rokhsar-Kivelson point~\cite{Horssen2015, Castelnovo2005}). %The peak height grows with $L$, from $5.8\times 10^{3}$ at $L = 10$ to $7.7\times 10^{4}$ at $L = 16$; with four system sizes and the resolution of the $s$-grid ($\delta s = 0.004$ around the peak), we do not extract its scaling.
At $s=0$, the Hamiltonian becomes 
\begin{align}
	\hat{H}_\mathrm{East}^{(s = 0)} = \frac{1}{2}(1-\hat{\sigma}_1^x) + \frac{1}{2}\sum_{j=1}^{L-1}\hat{n}_j(1-\hat{\sigma}_{j+1}^x),
\end{align}
a sum of positive semidefinite terms that all annihilate $\ket{+}^{\otimes L}$ such that the ground state is exactly $\ket{+}^{\otimes L}$ with zero energy.

For finite $s$, we obtain the ground state numerically using the wall-Chebyshev projector~\cite{Zhao2024}. For $s>0$, it is exponentially localized near the first site, with both $\mean{\hat n_j}$ and $\mean{\hat\sigma^x_j}$ decaying exponentially in $j$. For $s<0$, the corresponding localization lengths diverge and these observables become spatially homogeneous~\cite{Pancotti2020}. This delocalization-localization transition in real space also signals the QPT at $s=0$ [Fig.~\ref{fig:num_dnsty_grnd}].

Following our approach for $s\to-\infty$, we compute the fractal dimensions for various values of $s$, see Fig.~\ref{fig4}(b). We find that, in the $Z$-basis, $\gs$ is localised in Fock space for $s>0$  and delocalised for $s<0$ (multifractal in the $Z$ basis). On the other hand, in the $X$-basis it is delocalised for $s>0$, consistent with a product state in the $Z$-basis, while it is multifractal for $s<0$. The QPT is therefore also reflected in the fractal dimensions.

Seeking an approximation to the $\gs$ for arbitrary $s$ in the form of Eq.~\eqref{eq:def_spin-cohere}, we minimize its energy with $\theta_j=\theta$. For periodic boundary conditions, this is $\frac{L}{2}\cos^2\frac{\theta}{2} - \frac{e^{-s}}{4}L \sin\theta\del{1 + \cos\theta}$. 
The angle minimising this for each $s$ is shown in the inset to Fig.~\ref{fig1}(b), and the main panel shows the overlap of the ansatz at this angle with the exact $\gs$. Starting from $\frac{\pi}{3}$ for $s\to -\infty$, $\theta$ rotates to $\frac{\pi}{2}$ at $s=0$, with the rotated ansatz remaining a good approximation throughout. It also remains a good approximation for $s>0$ except for a small region just above $s=0$~\footnote{Just above $s=0$ the $\gs$ becomes exponentially localized near the first site, which a uniform product state cannot capture, and the overlap drops. For $s\gtrsim 0.2$ the optimal angle jumps to $\theta\approx\pi$, i.e.~the empty state $\ket{0}^{\otimes L}$, and the overlap recovers to $\approx 0.6$--$0.9$ for $L = 18$.}.

%================================
%	Peculiar excited states

\begin{figure}[t]
	\centering
	\includegraphics[width=\linewidth]{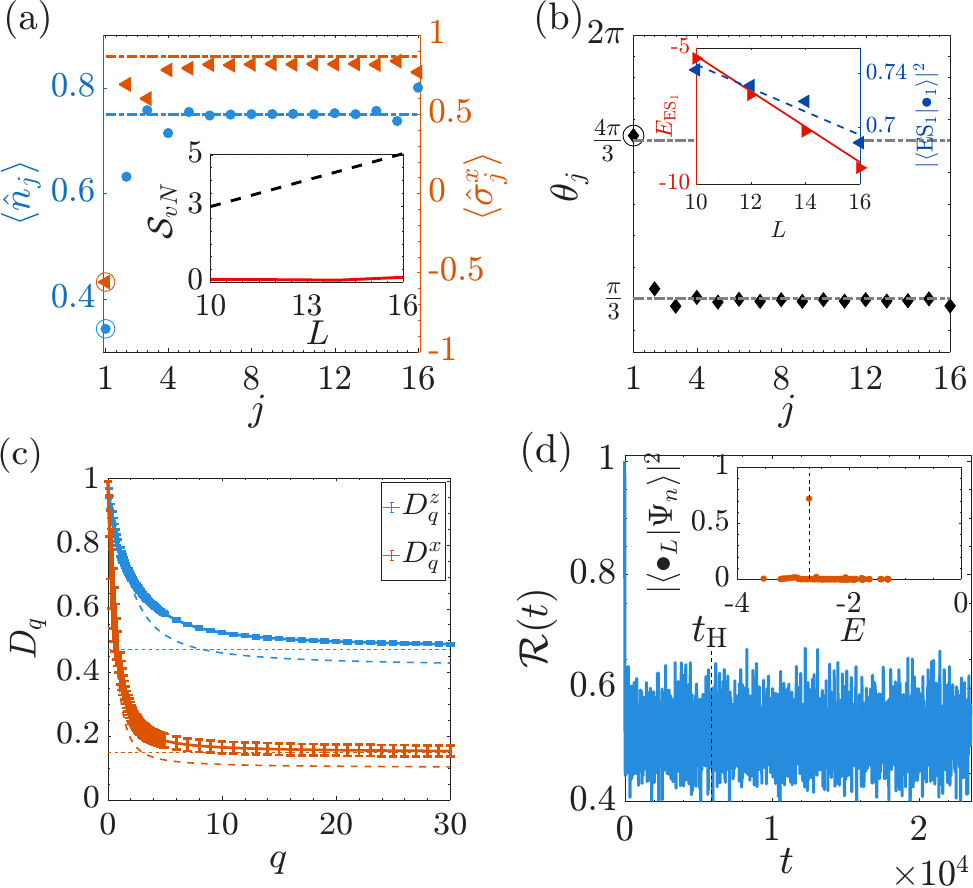}
	\caption{ All panels in this figure for $s\to-\infty$. (a)~Number density and magnetization along X-directions for $\esl$ [Eq.~\eqref{eq:spin-c_apprx_ES}] versus lattice site for $L = 16$. Horizontal dashed lines denote $\mean{\hat{n}} = \frac{3}{4}$ and $\mean{\hat{\sigma^x}} = \frac{\sqrt{3}}{2}$, as expected for the spin-polarized state $\ket{\frac{\pi}{3}}^{\otimes L}$. Note that the first spin deviates significantly from the dashed lines for both observables. Inset shows the entanglement entropy of $\esl$ vs.~system size where the dashed line denotes the Page values.
		%------------
		(b)~Optimum angles of a spin-coherent state [Eq.~\eqref{eq:def_spin-cohere}] maximizing the overlap with $\esl$. Inset shows the energy of the spin-coherent state $\ket{\bullet_1}$ (solid line) and $\esl$ (markers) on the left axis, and the overlap $|\mean{\mathrm{ES_1}|\bullet_1}|^2$ on a logarithmic scale on the right axis, as a function of system size; the dashed line is a linear fit, $\ln|\mean{\mathrm{ES_1}|\bullet_1}|^2 = -0.171 - 0.0121\,L$.
		%------------
		(c)~Fractal dimensions of $\esl$ in both Z- and X-bases. Horizontal dashed lines denote $D_\infty$. Error-bars denote 95\% confidence interval.
		%------------
		(d)~Return probability of the spin-coherent state. Vertical dashed line denotes the Heisenberg time. Inset shows the overlap of the spin-coherent state with the energy eigenstates where the vertical line denotes the energy of $\esl$.
	}
	\label{fig3}
\end{figure}
%================================

We now turn to the excited state $\esl$ with unusually low-entanglement compared to its neighboring eigenstates [Fig.~\ref{fig1}(a)]. In Fig.~\ref{fig3}(a), we show that for $s\to-\infty$, the expectation values of the local observables w.r.t.~$\esl$ are homogeneous over the real space except the first site. We also find that the entanglement entropy of $\esl$, $\mathcal{S}_\mathrm{vN} \approx 0.1090$ is approximately independent of system size and much smaller than the Page value~\cite{Page1993}.

The low entanglement entropy and difference of local observables on the first lattice site compared to the rest together again suggest an ansatz of the form of Eq.~\eqref{eq:def_spin-cohere}. We obtain the angles $\theta_j$ by maximising the overlap with $\esl$ and show the resulting $\theta_j$'s for $L=16$ in Fig.~\ref{fig3}(b). We find that $\theta_j \approx \frac{\pi}{3}$ for $j=2, \ldots, L$ while $\theta_1 \approx \frac{4 \pi}{3}$ so that the $\esl$ differs from the $\gs$ only at the first spin which is flipped by an angle of $\pi$.  Denoting $\ket{\circ} = \ket{\frac{\pi}{3}}$ and $\ket{\bullet} = \ket{\frac{4\pi}{3}}$,
\begin{align}
	\esl\approx \ket{\bullet}\otimes\ket{\circ}^{\otimes (L-1)} \equiv \ket{\bullet_1}
	\label{eq:spin-c_apprx_ES}
\end{align}
The energy of this state, $\frac{9\sqrt{3}}{8} - \frac{3}{4} - \frac{3\sqrt{3}L}{8}$, closely approximates the energy of the eigenstate $\esl$ for all system sizes, as shown in Fig.~\ref{fig3}(b); both lie a size-independent energy above the ground state with $E_{\mathrm{ES}_1}-E_\mathrm{GS} = 2.260$ for $L=8$--$15$, compared to $\frac{5\sqrt{3}}{4}\approx 2.165$ for the two ansatz states from Eqs.~\eqref{eq:spin-pol_GS} and \eqref{eq:spin-c_apprx_ES}.

We find that the overlap $|\mean{\mathrm{ES_1}|\bullet_1}|^2$ decays exponentially with system size (the overlap is $0.74$ at $L=10$ and $0.69$ at $L=16$) while the ratio $|\mean{\mathrm{ES_1}|\bullet_1}|^2/|\mean{\mathrm{GS}|\Theta}|^2\approx 0.79$--$0.81$ is nearly independent of $L$ for our accessible system sizes. Thus, the edge part of the ansatz is size-independent and the fidelity decay comes entirely from the bulk. As a result, the return probability starting from $\ket{\bullet_1}$ again does not decay even beyond the Heisenberg time for the accessible sizes, see Fig.~\ref{fig3}(d). The fractal dimensions of $\esl$ are well approximated by Eq.~\eqref{eq:Dq_spin_cohere} [shown in Fig.~\ref{fig3}(c)], as the fractal dimensions of our ansatz remain the same when approximating both $\gs$ and $\esl$. The chiral symmetry of $\hat{H}_{-\infty}+\frac{\mathbb{I}}{2}$ implies that
\begin{align}
	\bigotimes_{j=1}^{L-1} \hat{\sigma}_j^z \hat{\sigma}_L^y \ket{\bullet_1} \propto \ket{\frac{2\pi}{3}}\otimes\ket{\frac{5\pi}{3}}^{\otimes (L-2)}\otimes \ket{\bullet}
\end{align}
has exactly the same overlap with the excited eigenstate $\esu$, which has energy $-E_{\mathrm{ES}_1}-1 \sim \mathcal{O}(L)$.

%----------------------------------
%	Peculiar excited state as edge mode
%================================
\begin{figure}[t]
	\centering
	\includegraphics[width=\linewidth]{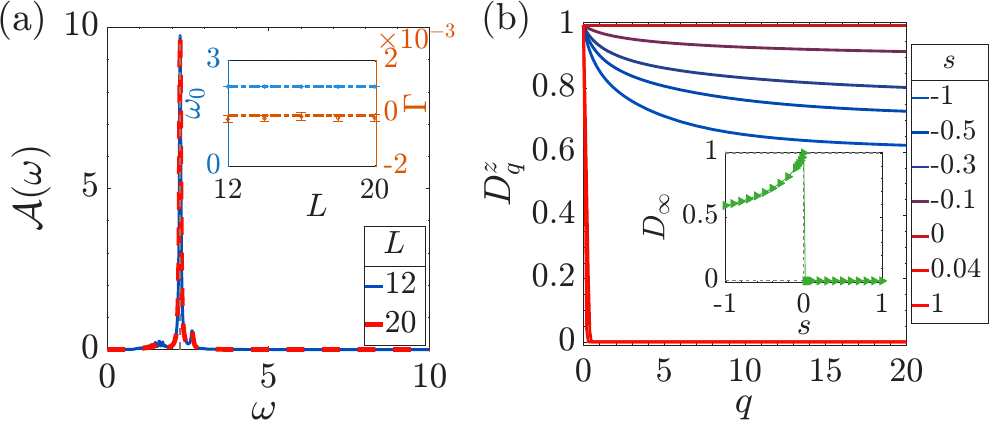}
	\caption{(a)~Spectral function (Eq.~(\ref{eq:A})) for a $\pi$-pulse on the first site of the actual ground state of $\hat{H}_{-\infty}$, for $L = 12$ and $20$ with broadening $\eta = 0.028$ (one Ritz-value spacing at the peak). Inset shows $\omega_0$ and $\Gamma$ (extrapolated to $\eta \to 0$) vs.~$L$; dash-dotted lines denote their $L\to\infty$ values.
		%--------------
		(b)~Fractal dimensions of the $\gs$ in Z-basis for different values of $s$. Inset shows $D_\infty$ vs.~$s$ capturing QPT at $s = 0$.
	}
	\label{fig4}
\end{figure}
%================================

Thus, we find that the excited but low-entanglement eigenstate $\esl$ embedded in a spectrum of highly-entangled eigenstates is close to the spin-coherent state $\ket{\bullet_1}$. This can be regarded as an edge mode~\cite{Kattel2025, Verresen2018, Asboth2016book, Kitaev2001, Hasan2010} of the East Hamiltonian at $s\to -\infty$. To establish this, we apply a local perturbation, $\hat{H} \to \hat{H} + \hat{\mathcal{O}}$ at the left edge of the lattice and look at the behavior of the modified low-energy states. As going from $\ket{\Theta}$ to $\ket{\bullet_1}$ requires applying a $\pi$-pulse on the first site around Y-axis (corresponding unitary transformation is $-i\hat{\sigma}_1^y$), we choose our local perturbation as $\hat{\mathcal{O}} = 2\hat{\sigma}_1^x - \hat{\sigma}_1^z\hat{\sigma}_2^x$~[Appendix~\ref{app:edge-mode}]. Then, the presence of an edge mode appears as a sharp peak in the spectral function of $-i\hat{\sigma}_1^y\gs$, which has an approximate Lorentzian form~\cite{Giamarchi2003book}
\begin{align}
	\mathcal{A}(\omega) \sim \frac{\mathcal{Z}}{\pi} \frac{\Gamma}{(\omega - \omega_0)^2 + \Gamma^2}
	\label{eq:A}
\end{align}
where $\omega_0$ is the energy, $\Gamma$ is the inverse lifetime, $\mathcal{Z}$ is the weight of the edge mode and $\gs$ is the ground state of $\hat{H}_{-\infty}$. 

In Fig.~\ref{fig4}(a), we show that the spectral function $\mathcal{A}(\omega)$ is sharply peaked at a size-independent energy $\omega_0 = 2.2598 \pm 0.0002$, equal to $E_{\mathrm{ES}_1}-E_\mathrm{GS} = 2.260$, with weight $\mathcal{Z} = 0.842 \pm 0.003$ and a width that vanishes within our resolution, $\Gamma = 0.0000 \pm 0.0003$ (Table~\ref{tab:spectral_fn}), while the spectral width of $\hat{H}_{-\infty}$ scales as $\sim \sqrt{L}$. % The peak is due to the single eigenstate, $\esl$: exact diagonalisation shows one eigenstate carrying weight $0.82$--$0.84$ at $\omega = 2.260$ for $L = 8$--$12$. 
Note that $-i\hat{\sigma}_1^y\ket{\Theta} = \ket{\bullet_1}$ up to a phase, so $|\mean{\bullet_1|(-i\hat{\sigma}_1^y)|\mathrm{GS}}|^2 = |\mean{\Theta|\mathrm{GS}}|^2$ exactly. Thus, the spectral function shows that $\ket{\bullet_1}$ is an edge mode with the excitation pinned to the first site having long lifetime despite the bulk eigenstates being ergodic. We note that this excitation lies only $\mathcal{O}(1)$ above the ground state, i.e.~at zero energy density; it is a low-energy edge excitation, which is what makes its long lifetime possible.%~[Appendix~\ref{app:edge-mode}].

%----------------------------------
Away from $s\to -\infty$, for $s\lessapprox -1$, the state $\ket{\bullet_1}$ still has strong overlap with a single eigenstate, while for $-1\lessapprox s < 0$ it develops appreciable overlap with two eigenstates with different energies, the gap between which is independent of the system size %(at $s=-0.2$, $0.04167$ to five digits for $L=8$--$13$) 
and decreases upon increasing $s$, vanishing at $s=0$ [Fig.~\ref{fig1}(c)]. The two overlaps decrease slowly with $L$ (Fig.~\ref{fig:esl_olap}), % (at $s=-0.2$, from $0.495$ to $0.401$ and from $0.300$ to $0.243$ between $L=6$ and $13$, i.e.~by $\approx 3\%$ per site; Fig.~\ref{fig:esl_olap}), 
so the oscillation amplitude of the return probability of $\ket{\bullet_1}$ also decreases with $L$ while the oscillation frequency remains system size independent. As the gap between the overlapping eigenstates decrease with $s$, the oscillation time period of the dynamics of $\ket{\bullet_1}$ increases as we approach $s = 0$.

Thus, for $-1\lessapprox s < 0$, the locally-correlated initial state $\ket{\bullet_1}$ has significant overlap with two eigenstates and results in long-lived oscillations at the size-independent frequency set by their gap. We show this in Fig.~\ref{fig1}(d), where the time evolution of both the return probability (a global quantity) and $\mean{\hat{\sigma}^x_1(t)}$, the magnetization of the first spin along X-axis are shown. Since the excitation is localised at the edge, its contribution to the spatial average is $\mathcal{O}(1/L)$; the oscillation is robust in local observables near the edge whose Fourier amplitude extrapolates to a finite value [Fig.~\ref{fig:dynamics_all}]. Note that in interacting many-body systems, the eigenstates at the spectral edge can be athermal despite thermalization in the middle of the spectrum; however, this usually leads to equilibration of the reduced density matrix to the generalized Gibbs ensemble in case of low-energy initial states, not persistent oscillations as in our case. Let us remark here that the athermal initial state $\ket{\bullet_1}$ is not embedded in the bulk spectrum, but rather remains close (but above and distinct from) the $\gs$. The dynamics of the model in Fock space also does not resemble that in a hypercube-like Fock space, as is the case for the PXP model~\cite{Turner2018, Bhore2023, Perciavalle2025Arxiv, Perciavalle2026}. The oscillations here therefore do not arise from the hypercube-like Fock-space structure behind weak ergodicity breaking in the PXP model, forming a distinct class. They also do not appear to be related to those of quantum scars in the semiclassical sense~\cite{Pizzi2025}.

%----------------------------------
%	Spin-coherent approximations of the special eigenstates for s → -∞
%================================
\begin{figure}[t]
	\centering
	\includegraphics[width=0.8\linewidth]{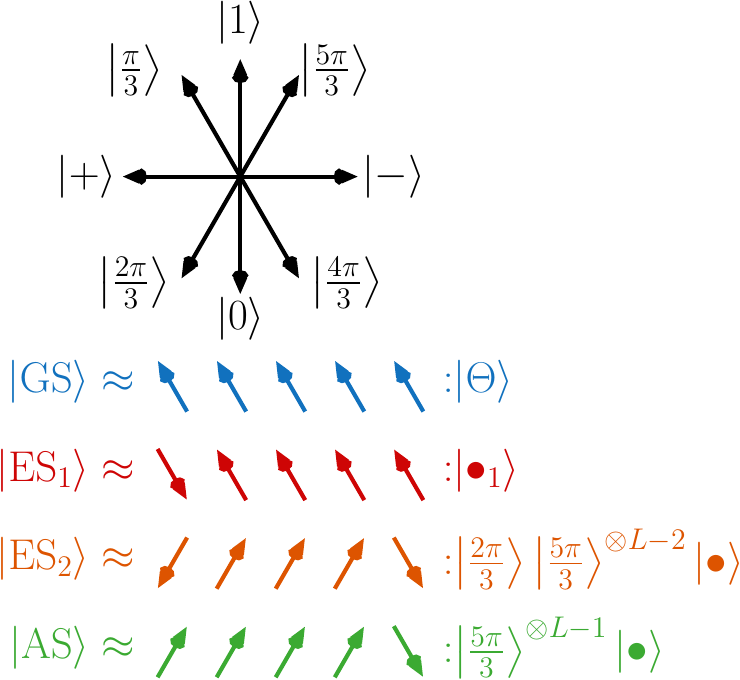}
	\caption{Spin-coherent approximations of the special eigenstates of the East Hamiltonian at $s \to -\infty$.
	}
	\label{fig5}
\end{figure}
%================================

%=============================
\section{Discussion}\label{sec_conclusion}
%=============================
We have studied the ground state and boundary-induced coherent dynamics of the quantum East model with open boundary condition and no global conserved charge. In the limit $s\to-\infty$, a spin-polarized state provides an accurate approximation to the ground state, capturing its local observables and the multifractal structure in the Fock space. For finite $s<0$, the optimal spin orientation changes continuously ($\frac{\pi}{3} \leq \theta \leq \frac{\pi}{2}$) towards the exact product ground state at $s=0$. The fidelity susceptibility, local observables and fractal dimensions provide complementary signatures of the quantum phase transition at this point.

We identified an unusually low-entanglement excited eigenstate, $\esl$, [see Fig.~\ref{fig5}] whose spin-coherent approximation $\ket{\bullet_1}$ differs from that of the ground state by a $\pi$-rotation of the first spin. Its excitation energy above the ground state is independent of system size, and the overlap of this simple boundary-excited product state with $\esl$, relative to the overlap of $\ket{\Theta}$ with the ground state, is independent of $L$. The sharp spectral response to a boundary perturbation, with an approximately size-independent excitation energy and a linewidth smaller than that energy, further supports the interpretation of $\ket{\bullet_1}$ as a long-lived boundary excitation.

At finite negative $s$ closer to the transition, the boundary-excited product state acquires appreciable overlap with two eigenstates. Their energy separation is approximately independent of system size and decreases towards zero as $s\to0$. This produces long-lived oscillations at a size-independent frequency in the non-equilibrium dynamics near the boundary, most clearly in local observables close to the edge; the overlaps with the two eigenstates, and hence the oscillation amplitude of the return probability, decrease slowly with $L$. This mechanism does not rely on the hypercube-like Fock-space dynamics associated with the PXP model, and the oscillating state lies only $\mathcal{O}(1)$ above the ground state. Our results therefore identify boundary physics as a route to coherent dynamics in a kinetically constrained system which still thermalizes at infinite temperature.

There are several interesting questions to follow up on. Are there other families of states analogous to \(\esl\) that yield long-lived oscillations for suitable initial conditions? Can the mechanism identified here be placed in a broader ``scar'' framework, either to scars as in~\cite{Turner2018} or to the many-body version of the semiclassical scar phenomenology as discussed in Ref.~\cite{Pizzi2025}, or is our $\esl$ in a fundamentally distinct class? How universal is the phenomenon across kinetically constrained models such as Fredrickson-Anderson-type models, and how does it depend on the choice of boundary terms? Finally, how robust are these oscillations to changes in the microscopic dynamics--can they survive in a Floquet/circuit implementation of the constraint, or under Trotterization of Eq.~\eqref{eq:H_East_EBC}?

%=============================
%	---ACKNOWLEDGEMENT---
%=============================
\begin{acknowledgments}
We thank J.~P.~Garrahan, I.~M.~Khaymovich and Andrea Pizzi for many useful discussions. We acknowledge support from the Leverhulme Trust Research Project Grant RPG-2025-063. 
\end{acknowledgments}

%=============================
%	---APPENDIX---
\appendix
%\renewcommand{\theequation}{A\arabic{equation}}
%=============================
%=============================
\section{Density of states}\label{app:DOS}
%=============================

\begin{figure}[t!]
	\centering
	\includegraphics[width=\linewidth]{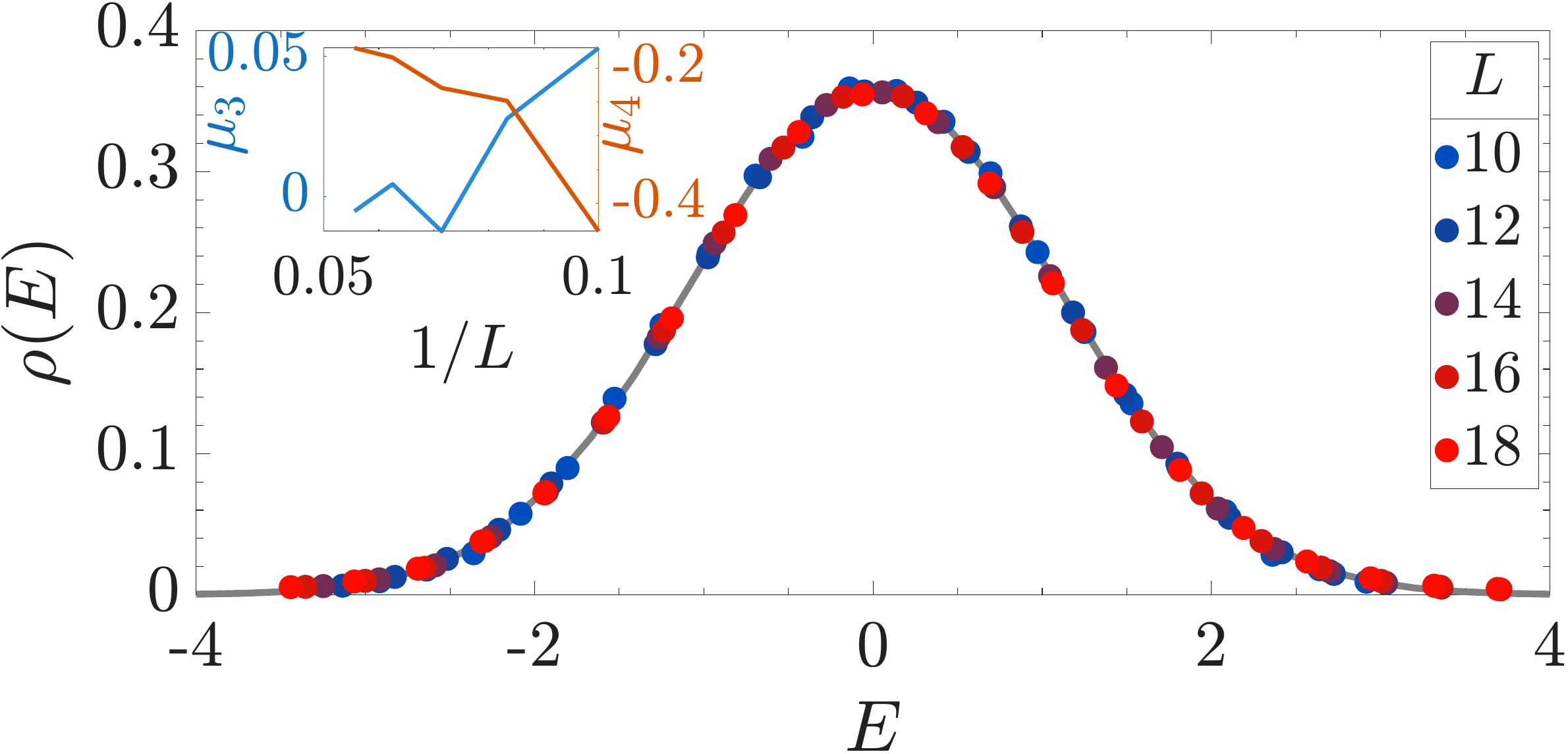}
	\caption{\textbf{DOS for $\hat{H}_\mathrm{East}(s\to-\infty)$:} DOS for different system sizes. The energy axis is centered and scaled to ensure zero mean and unit variance. Grey solid line denotes Gaussian fit. Inset shows the measures of skewness ($\mu_3$) and excess kurtosis ($\mu_4$) as a function of $L^{-1}$.}
	\label{fig:East_L_DOS}
\end{figure}

In the Z-basis, the East Hamiltonian in the limit $s\to-\infty$ has diagonal contributions only from $-\hat{n}_L$, so, $2^{L-1}$ number of Fock states with $L$th spin being up have energy $-1$. Then, $\mathrm{Tr}{H} = 2^{L-1} \times (-1)$. For $\hat{H}^2$, the diagonal contributions come from $\mathbb{I} + \sum_{j = 1}^{L} \hat{n}_j^2$ as $(\hat{\sigma}^x)^2 \equiv \mathbb{I}$ such that $\mathrm{Tr}{\hat{H}^2} = \mathrm{Tr}{ \mathbb{I} + \sum_{j = 1}^{L} \hat{n}_j^2 } = 2^L + L\times 2^{L-1}$. Then, the 1st and 2nd moments of energy are
\begin{align}
	\mean{E} = - \frac{1}{2}, \quad \mean{E^2} = 1 + \frac{L}{2}.
\end{align}
The density of states (DOS) for different system sizes collapse upon the scaling $E \to \frac{E + \frac{1}{2}}{\sqrt{\frac{L}{2} + \frac{3}{4}}}$. We find that the DOS is well approximated by a Gaussian distribution (Fig.~\ref{fig:East_L_DOS}).

%================================
%   s → -∞ limit
%================================

%================================
\section{Fractal dimensions}
\label{app:fractal-dims-limit}
%================================
To obtain the fractal dimensions, we start from the inverse participation ratios
\begin{align}
	I_q^{(N)} = \sum_{n = 1}^N |\Psi(n)|^{2q} \sim N^{(1-q)D_q^{(N)}}.
\end{align}
By extrapolating $D_q^{(N)}$ to infinite system size (assuming that the finite size corrections are polynomial functions of $1/L$~\cite{Luitz2020, Das2024}), we extract the fractal dimensions, $D_q$. 

In the Z-basis, the spin-polarized state $\ket{\Theta} = 2^{-L} \begin{pmatrix}
	1\\
	\sqrt{3}
\end{pmatrix}^{\otimes L}$ such that the IPRs become
\begin{align}
	\begin{split}
		I_q^{(N)} &= 2^{-2qL}(1 + 3^q)^L \Rightarrow D_q^z = \frac{\log_2(1+3^q) - 2q}{1 - q}
	\end{split}
	\label{eq:Dq_Z_spin_cohere}
\end{align}
Then, for $q\gg 1$, the fractal dimensions of the spin-coherent state can be written as $(\log_2 3 - 2)\frac{q}{1-q}$. In particular, the maximum intensity scales as $\frac{3^L}{2^{2L}}$ such that $D_\infty^z = 2 - \log_2 3 \approx 0.41$. Numerically, we find that $D_\infty^z \approx 0.47$ for the actual ground state.

%================================
%   s → -∞ limit, ground state
\begin{figure}[t!]
	\centering
	\includegraphics[width=\linewidth]{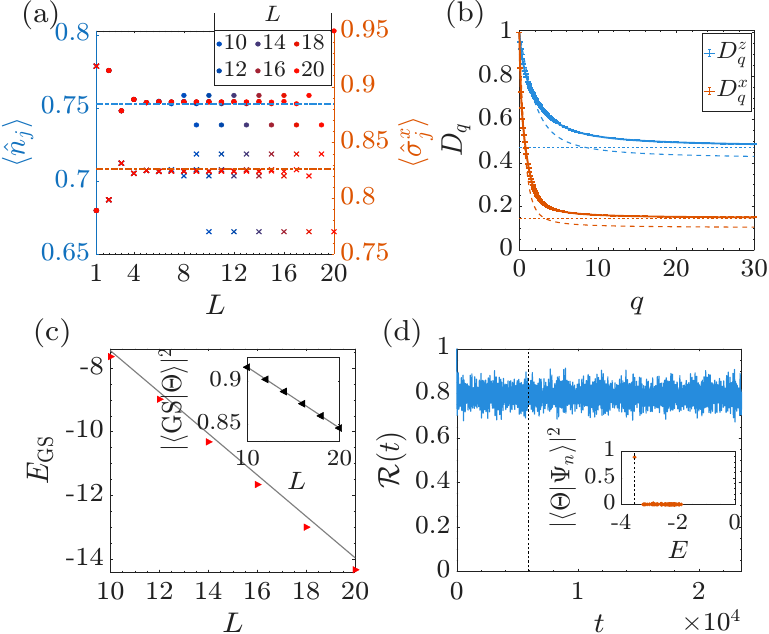}
	\caption{\textbf{Ground state for $\hat{H}_\mathrm{East}(s\to-\infty)$:} (a)~Number density (circle) and magnetization in X-direction (cross) w.r.t.~ground state for different system sizes shown via different colors. Dashed lines denote the average over the sites and different system sizes.
		%==========================
		(b)~Fractal dimensions of the ground state w.r.t.~Z and X bases. Horizontal dotted lines denote $D_\infty$ extracted from the system size scaling of the maximum intensity. Dashed lines denote the fractal dimensions of the spin-polarized state for $\theta = \frac{\pi}{3}$ (Eqs.~\eqref{eq:Dq_Z_spin_cohere} and \eqref{eq:Dq_X_spin_cohere}).
		%==========================
		(c)~Ground state energy vs.~system size via markers while the solid gray line shows the energy of the spin-polarized state, $\ket{\Theta}$. Inset shows the overlap $|\mean{\mathrm{GS}|\Theta}|^2$ as a function of $L$ on a logarithmic scale; the solid line is a linear fit, $\ln|\mean{\mathrm{GS}|\Theta}|^2 = -0.0036 - 0.0083\,L$, i.e.~the overlap decreases by $\approx 0.8\%$ per site.
		%==========================
		(d)~Return probability of $\ket{\Theta}$. Vertical dashed line denotes the Heisenberg time. Inset shows the overlap of $\ket{\Theta}$ with the energy eigenstates where the vertical dashed line denotes the ground state energy. We only show the energy levels for which the overlap is more than $N^{-1}$.
	}
	\label{fig:East_L_ground}
\end{figure}
%================================

Following the same procedure as above, we can calculate the fractal dimensions of the spin-coherent state in the X-basis, where $\ket{\Theta} = \del{\frac{\sqrt{3} + 1}{\sqrt{8}}}^L \begin{pmatrix}
	1\\
	\sqrt{3} - 2
\end{pmatrix}^{\otimes L}$. The corresponding IPRs are
\begin{align}
	\begin{split}
		I_q^{(N)} &= \del{\frac{\sqrt{3} + 1}{\sqrt{8}}}^{2qL} \del{1 + \del{2 - \sqrt{3}}^{2q}}^L\\
		\Rightarrow D_q^x &= \frac{2\del{\log_2(\sqrt{3}+1) - 3/2}q + \log_2(1 + (\sqrt{3} - 2)^{2q})}{1-q}
	\end{split}
	\label{eq:Dq_X_spin_cohere}
\end{align}
Particularly, the maximum intensity in the X-basis scales as $\frac{\del{\sqrt{3} + 1}^{2L}}{8^L}$ such that $D_\infty^x = 3 - 2\log_2\del{\sqrt{3}+1} \approx 0.10$. Numerically we find that $D_\infty^x \approx 0.14$ for the ground state. In Fig.~\ref{fig:East_L_ground}(b), we show the analytical expressions of the fractal dimensions of the spin-coherent state along with those of the ground state in both Z- and X-bases. 

%================================
\section{Ground state localisation for finite $s$}
\label{app:gs-loc}
%================================
%   Number density w.r.t. ground state
\begin{figure}[t]
	\centering
	\includegraphics[width=\linewidth]{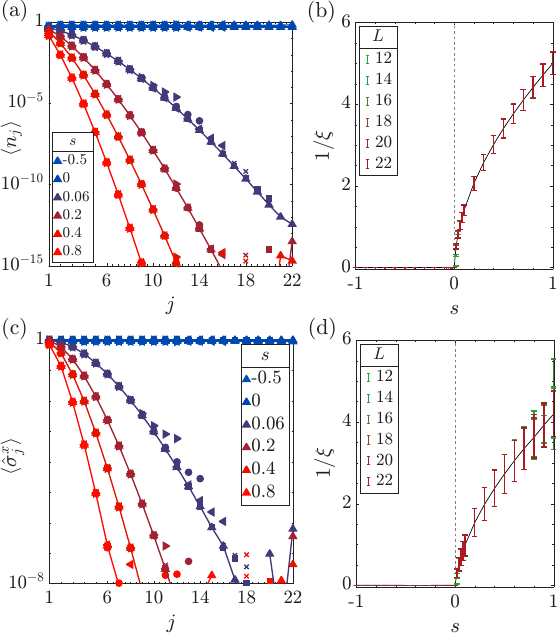}
	\caption{\textbf{Local observables w.r.t.~the ground state:} (a)~Number density w.r.t.~the ground state for different values of $s$. Markers denote different system sizes, $L = 12, 14,...,22$.
		%=======================
		(b)~The localization length extracted from the exponential decay of the number density w.r.t.~lattice sites.
		%=======================
		(c)~Magnetization in the X-direction averaged over the ground state for different values of $s$.
		%=======================
		(d)~The localization length extracted from the decay of the X-component of the magnetization. Error-bars denote 95\% confidence interval.
	}
	\label{fig:num_dnsty_grnd}
\end{figure}
%================================

The single-site occupation $\mean{\hat{n}_j} \propto \exp\del{-\frac{j}{\xi}}$ where $\xi$ is the real-space localization length. For $s<0$, $\mean{\hat{n}_j}$  in \gs is a homogeneous function of $j$~\cite{Pancotti2020}. $\xi$ as extracted from the decay of $\mean{\hat{n}_j}$ diverges at $s = 0$ approximately as $\xi \sim s^{-\frac{1}{2}}$ (a single-size fit for $L = 22$ gives the exponent $0.5481 \pm 0.0008$; the error is statistical only and does not include finite-size effects).

%================================
%   Fractal dimensions of ground state
\begin{figure}[t]
	\centering
	\includegraphics[width=\linewidth]{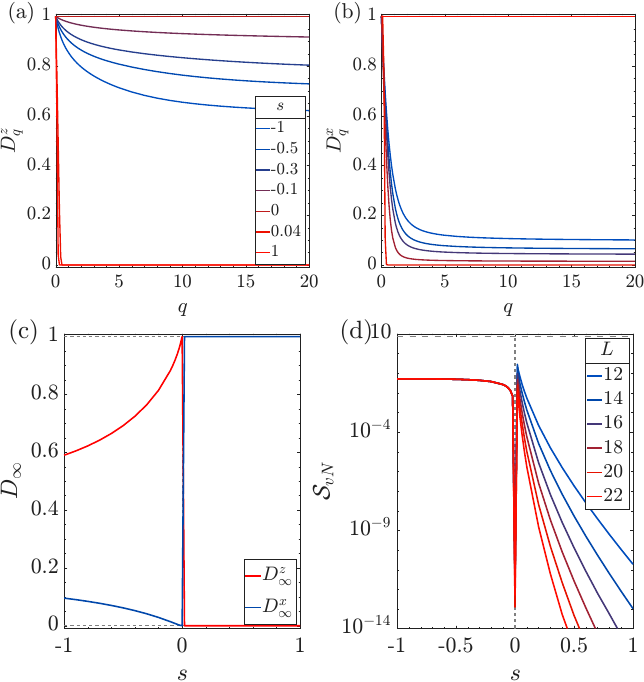}
	\caption{\textbf{Fractal dimensions of the ground state:} Fractal dimensions w.r.t.~(a)~Z-basis and (b)~X-basis for various values of $s$.
		%===========================
		(c)~$D_\infty$ obtained from the system size scaling of the maximum intensity in both Z- and X-bases.
		%===========================
		(d)~Bipartite von Neumann entanglement entropy of the ground state for different system sizes. The horizontal dashed line denotes the Page value for $L = 20$.
	}
	\label{fig:Dq_grnd}
\end{figure}
%================================

We also compute the expectation value of the magnetization in X-direction, $\mean{\hat{\sigma}^x_j}$ w.r.t.~the ground state at each lattice site. We observe a similar behavior as that of the number density: for $s < 0$, $\mean{\hat{\sigma}^x_j}$ is homogeneously spread over all the lattice sites while for $s>0$, an exponential decay from the first lattice site is observed, $\mean{\hat{\sigma}^x_j} \propto \exp\del{-\frac{j}{\xi}}$. For $s = 0$, $\mean{\hat{\sigma}^x_j} = 1$ for all values of $j$ as the ground state becomes $\ket{+}^{\otimes L}$. The localization lengths extracted from the exponential decay show power-law divergence at $s = 0$ (a single-size fit for $L = 22$ gives the exponent $0.6198 \pm 0.0047$; this differs from the value obtained from $\mean{\hat n_j}$ by more than the statistical errors, so finite-size effects are not negligible and we do not quote a common exponent).

We find that for $s>0$, $D_q^z = 0$ for any $q > 0$, which implies that the ground state is localized in the Fock space as well as in the real space (reflected in $\mean{\hat{n}_j}$ vs.~$j$). On the other hand, for $s<0$, we find that $0 < D_q^z < 1$ for all values of $q > 0$ and $D_q^z > D_{q'}^z$ for $q < q'$. Thus, the ground state of the East model is multifractal over the Z-basis for $s < 0$, despite being homogeneously spread over the real space.

As opposed to the Z-basis, we find the ground state to be uniformly spread over all the X-basis states for $s > 0$, as reflected by $D_q^x = 1$ for all values of $q$. This is expected as a product state in Z-basis is completely extended in the X-basis. On the other hand, the ground state is multifractal in X-basis for $s < 0$ where the fractal dimensions are bounded by $0 < D_\infty^x < 1$. Thus, by looking at the fractal dimensions of the ground state, we are able to identify the QPT at $s = 0$.

We find that $D_\infty^x \to 0$ and $D_\infty^z \to 1$ for $s \to 0$, as shown in Fig.~\ref{fig:Dq_grnd}(c). Previously, we saw that $\mean{\hat{\sigma}_j^x} = 1$ w.r.t.~the ground state for all the lattice sites at $s = 0$. This is consistent with the exact ground state $\ket{+}^{\otimes L}$ at $s = 0$ (Sec.~\ref{sec_finite_s}), which is completely extended (localized) in the Z-basis (X-basis). This is also reflected in the bipartite von Neumann entanglement entropy of the ground state, which goes to zero at $s = 0$ and exhibits a discontinuity, as shown in Fig.~\ref{fig:Dq_grnd}(d). For all values of $s$, the entanglement entropy of the ground state remains much smaller than the Page value observed in case of the maximally entangled states~\cite{Page1993}.

%--------------------------------
%   Anti-ground state
\section{Anti-ground state for $s\to-\infty$}
\label{app:as}
%================================
%   s → -∞ limit, anti-ground state
\begin{figure}[t]
	\centering
	\includegraphics[width=\linewidth]{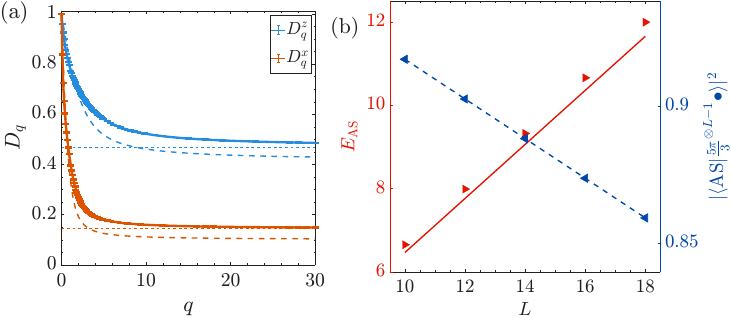}
	\caption{\textbf{Anti-ground state for $\hat{H}_\mathrm{East}(s\to-\infty)$:} % (a)~Site density and magnetization in X-direction w.r.t.~anti-ground state (asymptotic values are 0.754 and -0.8243, respectively). Solid lines denote cubic fitting w.r.t.~$L^{-1}$.
		%==========================
		(a)~Fractal dimensions of the anti-ground state w.r.t.~Z and X bases. Horizontal dotted lines denote $D_\infty$ extracted from the system size scaling of the maximum intensity. Dashed lines denote the analytical expressions from Eqs.~\eqref{eq:Dq_Z_spin_cohere} and \eqref{eq:Dq_X_spin_cohere}.
		%==========================
		(b)~Anti-ground state energy (left axis) as a function of system size via markers while the solid line shows the energy of the ansatz in Eq.~\eqref{Eq:AS_ansatz}, $-E_\Theta-1 = \frac{\sqrt{3}(3L+1)}{8}-\frac{1}{4}$. The right axis shows the overlap between the anti-ground state and the ansatz as a function of $L$ on a logarithmic scale, equal to $|\mean{\mathrm{GS}|\Theta}|^2$ by chiral symmetry; the dashed line is a linear fit, $\ln|\braket{\mathrm{AS}|\frac{5\pi}{3}}^{\otimes L-1}\ket{\bullet}|^2 = -0.0036 - 0.0083\,L$.
	}
	\label{fig:East_L_anti-ground}
\end{figure}

From the chiral symmetry of $\hat{H}_{-\infty} + \frac{\mathbb{I}}{2}$, we propose an ansatz [Eq.~\eqref{Eq:AS_ansatz}] for the anti-ground state. The fractal dimensions of the ansatz follow Eqs.~\eqref{eq:Dq_Z_spin_cohere} and \eqref{eq:Dq_X_spin_cohere}. In Fig.~\ref{fig:East_L_anti-ground}(a), we show the fractal dimensions of the anti-ground state along with those of the spin-coherent ansatz and find a nice agreement. Thus, the anti-ground state is multifractal in both Z- and X-bases, similar to the ground state. In Fig.~\ref{fig:East_L_anti-ground}(b), we show the agreement between the energies of the actual anti-ground state and the spin-coherent ansatz as a function of system size. By the chiral symmetry, the energy of the ansatz is $-E_\Theta-1$ and its overlap with the anti-ground state equals $|\mean{\mathrm{GS}|\Theta}|^2$ for every $L$, shown on the right axis.

%================================
\section{Energy correlations for $s\to-\infty$}
\label{app:energy-corr}
\begin{figure}[t!]
	\centering
	\includegraphics[width=\linewidth]{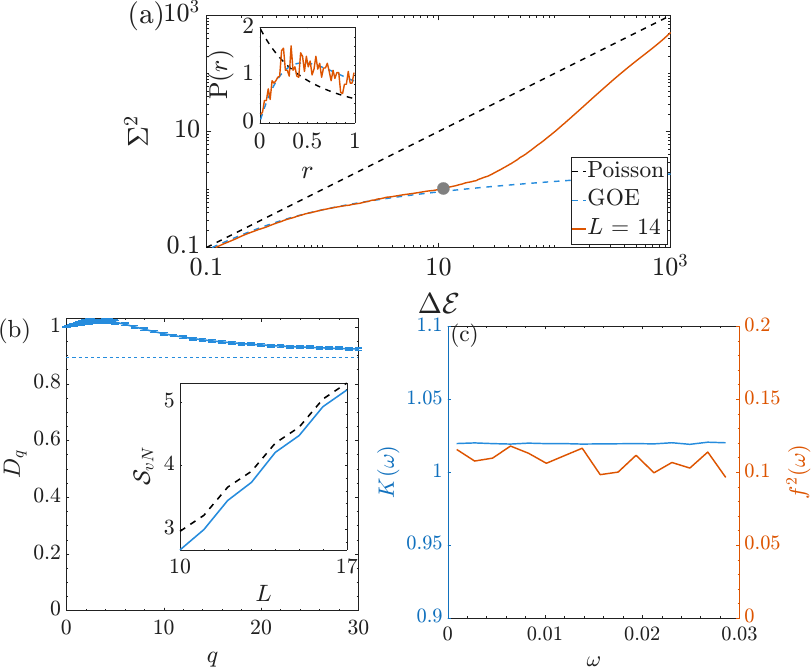}
	\caption{\textbf{$\hat{H}_\mathrm{East}(s\to-\infty)$:} (a)~Number variance for $L = 14$ averaged over energy windows with centers in the middle 50\% of the spectrum. Dashed lines denote the analytical expressions of the Poisson and GOE. Marker denotes the Thouless energy. Inset shows the density of the ratio of level spacing along with the analytical expressions of the Poisson and GOE.
		%------------------
		(b)~Fractal dimensions averaged over middle 256 eigenstates for $L = 17$. Inset shows the entanglement entropy averaged over middle 256 eigenstates as a function of system size. Black dashed line denotes the Page values.
		%------------------
		(c)~Overlap of the eigenstates and spectral function of $\hat{n}_{L/2}$ vs.~energy gap for $L = 16$.
	}
	\label{fig:East_L_Thouless}
\end{figure}

%================================
Above analyses raise the question: are all the bulk states (away from the spectral edges containing $\mathcal{O}(1)$ states) ergodic? If yes, then all such bulk states strongly hybridize with each other and leads to correlated energy levels in the middle of the spectrum. Thus, to probe ergodicity in our system, first, we look at the energy correlations. To quantify the nature of short-range correlations (on the scale of mean level spacing), we look at the ratio of level spacing~\cite{Oganesyan2007, Atas2013, Das2019, Das2026}. The average ratio of level spacing is approximately 0.53 (0.39) for the GOE (Poisson ensemble). We find that for the East Hamiltonian at $s \to -\infty$, the average ratio of level spacing is approximately 0.53 and corresponding distribution follows the analytical expression of that of the GOE, as shown in the inset of Fig.~\ref{fig:East_L_Thouless}(a). Thus, nearby bulk energy levels are strongly correlated as in random matrix theory.

%================================
%   s → -∞ limit, Thouless energy
%================================

%================================
%   Long-range energy correlation
The presence of short-range correlation however does not guarantee that two energy levels with gap much larger than the mean level spacing are correlated, which is expected in an ergodic system. To probe such long-range correlation, we look at the level number variance~\cite{Mehta2004book, Das2025a, Das2026, Roy2025}. In case of GOE, where all the energy levels are correlated, the number variance exhibits a logarithmic behavior. Contrarily, the number variance is linear in case of the Poisson ensemble, where any two energy levels are uncorrelated. In physical systems, there may exist a mesoscopic energy scale called the Thouless energy, which is typically smaller (larger) than the global energy bandwidth (mean level spacing)~\cite{Serbyn2017, Sierant2020, Corps2020, Das2022a, Das2025, Das2025a, Das2022b, Seth2026}. Energy levels with gap smaller than the Thouless energy are correlated similar to the random matrix models whereas energy correlations cease to exist on a larger energy scale.

To compute the number variance, it is necessary to unfold the energy spectrum to get rid of the model dependent global trend of the DOS such that mean level spacing becomes unity~\cite{Guhr1998}. We denote the unfolded energy levels as $\Ecal$. In Fig.~\ref{fig:East_L_Thouless}(a), we show the level number variance of the East model in the $s\to -\infty$ limit averaged over energy windows of length $\Delta \Ecal$ and centers over the middle 50\% of the spectrum. We identify the Thouless energy as the energy scale above which the number variance deviates from that of the GOE. For $L = 14$, the Thouless energy is $\sim\mathcal{O}(10)$, which is larger than the mean level spacing ($\mathcal{O}(1)$) but much smaller than the global bandwidth ($\mathcal{O}(N)$). This is generic for ergodic systems with local interactions, where levels separated by more than the Thouless energy are not correlated as in random matrix theory.

%================================
%   Dynamics
\begin{figure*}
	\centering
	\includegraphics[width=\linewidth]{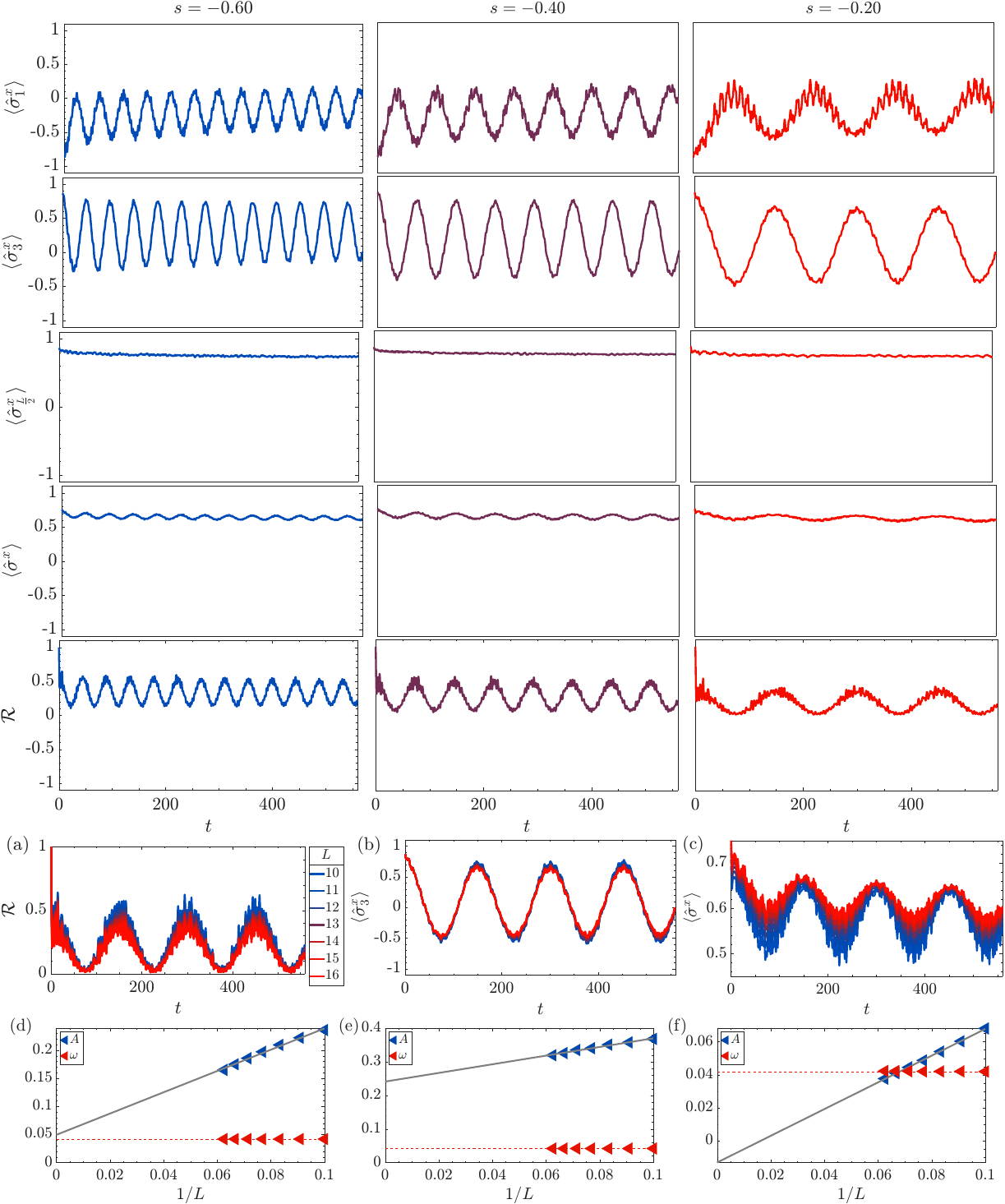}
	\caption{ Dynamics of the East model starting from $\ket{\bullet_1}$. For $L = 16$ and different values of $s$, the rows correspond to the local magnetization in the X-direction at the first site ($\langle \hat{\sigma}_1^x\rangle$), a fixed distance away from the boundary ($\langle \hat{\sigma}_{3}^x\rangle$), bulk spin ($\langle \hat{\sigma}_{\frac{L}{2}}^x\rangle$), spatial average ($\langle \hat{\sigma}^x\rangle \equiv \frac{1}{L}\sum_j \langle \hat{\sigma}_j^x\rangle$) and the return probability ($\mathcal{R}$). Panels (a), (b) and (c) respectively show the return probability, local magnetization in the X-direction of the third spin and the spatially averaged magnetization for different system sizes and $s = -0.2$. Amplitude ($A$) of the dominant mode (other than DC shift) of the Fourier transform of the different dynamical observables is shown in panels (d), (e) and (f) as a function of the inverse system size for $s = -0.2$. The amplitude for $\langle \hat{\sigma}_{3}^x\rangle$ extrapolates to $0.24 \pm 0.01$ (solid line: linear fit in $L^{-1}$), while those of $\mathcal{R}$ and of the spatial average decrease with $L$ (extrapolating to $0.05 \pm 0.02$ and $\approx 0$, respectively), as expected for an edge excitation. The frequency $\omega$ of the dominant mode, obtained from the zero-padded Fourier transform, is $0.0418$ for all $L$, equal to the size-independent gap $|E_{n_1}-E_{n_2}| = 0.04167$ (dashed lines; period $2\pi/0.04167\approx 151$). Time and frequency are in units of Eq.~\eqref{eq:H_East_EBC}.
	}
	\label{fig:dynamics_all}
\end{figure*}
%================================

To understand the manifestation of the Thouless energy in the middle of the spectrum, we obtain middle 256 eigenstates for different system sizes. In Fig.~\ref{fig:East_L_Thouless}(b), we show the fractal dimensions in the Z-basis averaged over the middle 256 eigenstates. We find that $D_q\approx 1$ for all values of $q$ as expected in ergodic states. This is also reflected in the von Neumann entanglement entropy, which exhibits volume law growth, as shown in the inset of Fig.~\ref{fig:East_L_Thouless}(b). To capture the degree of hybridization among the eigenstates, we look at the overlap~\cite{Kravtsov2015}
\begin{align}
	K(\omega) = N\sum_{n = 1}^{N} |\Psi_E(n)|^2 |\Psi_{E+\omega}(n)|^2
\end{align}
where $N = 2^L$ is the Hilbert space dimension, $\ket{\Psi_E}$ is the eigenstate with energy $E$. $K(\omega)\approx 1$ for two eigenstates with strong overlap as in the ergodic systems whereas a power-law decay is observed for $\omega > \Eth$. In Fig.~\ref{fig:East_L_Thouless}(c), we show the overlap as a function of energy gap for middle 256 eigenstates. We find that $K(\omega)\approx 1$, i.e.~all such states are strongly hybridizing. To support this conclusion, we also look at the spectral function of a local observable~\cite{Serbyn2017}
\begin{align}
	f^2(\omega) = e^{\shn(E)}\sum_{m, n} |\mathcal{O}_{mn}|^2 \delta\del{\omega - |E_m - E_n|}
\end{align}
where $\shn(E)$ is the thermodynamic entropy at $E \equiv (E_m + E_n)/2$ and $\mathcal{O}_{mn} \equiv \bra{\Psi_{E_m}} \hat{\mathcal{O}} \ket{\Psi_{E_n}}$. We consider the entropy to be $\shn(E) = \ln(N \rho(E))$, where $\rho(E)$ is the DOS and take the local observable to be $\hat{n}_{L/2}$. In Fig.~\ref{fig:East_L_Thouless}(c), we show that the spectral function is also homogeneous for the middle 256 states, $f^2(\omega)\approx 0.1$ without any structure over the energy window. Thus, the eigenstates in the middle of the spectrum typically behave as ergodic wavefunctions.

%================================
\section{Spectral function for the boundary excitation}
\label{app:edge-mode}
%================================
%   Spectral function
\begin{figure}[t!]
	\centering
	\includegraphics[width=\linewidth]{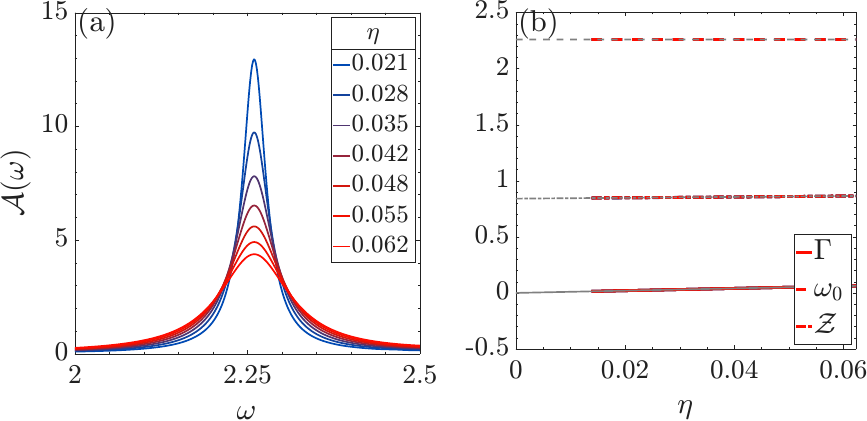}
	\caption{(a)~Spectral function for the boundary excitation (the $\pi$-pulse on the first site) in case of $\hat{H}_\mathrm{East}(s\to-\infty)$ for $L = 20$ and different values of the broadening parameter, $\eta$, from $0.5$ to $2.25$ times the spacing of the Ritz values at the peak ($0.028$). (b)~Inverse lifetime ($\Gamma$; solid lines), energy ($\omega_0$; dashed lines), and the weight of the quasiparticle peak ($\mathcal{Z}$; dot-dashed lines) as a function of $\eta$ for system sizes $L = 12, 14,\dots, 20$ (denoted by different colors). The gray lines denote the linear extrapolation to $\eta \to 0$; $\Gamma = \eta$ within the fit accuracy, i.e.~the intrinsic width vanishes.
	}
	\label{fig:East_L_spectral}
\end{figure}
%================================

To probe any coherent boundary excitation in the limit $s\to-\infty$, we apply a $\pi$-pulse on the first site such that the first qubit rotates by an angle $\pi$ around the Y-axis. The corresponding rotation operator is $\exp\del{-i\frac{\theta}{2}\hat{\sigma}_1^y}$ which reduces to $-i\hat{\sigma}_1^y$ for $\theta = \pi$.

For a Hamiltonian $\hat{H}$ with eigenbasis $\{ E_n, \ket{\Psi_n} \}$ and a unitary transformation $\hat{U}$, $\hat{U}\ket{\Psi_n}$ is an eigenstate of $\hat{U} \hat{H} \hat{U}^\dagger$ with energy $E_n$. Thus, a perturbation $\hat{U} \hat{H} \hat{U}^\dagger - \hat{H}$ to the original Hamiltonian changes the eigenstates as $\ket{\Psi_n} \to \hat{U}\ket{\Psi_n}$. In our case, the unitary transformation is $-i\hat{\sigma}_1^y$ and the Hamiltonian is $\hat{H}_{-\infty} = -\hat{\sigma}_1^x - \sum_{j = 1}^{L-1} \hat{n}_j\hat{\sigma}_{j+1}^x - \hat{n}_L$. Note that,
\begin{align}
	\begin{split}
		& (-i\hat{\sigma}_1^y) \hat{H}_{-\infty} (i\hat{\sigma}_1^y)\\
		= &-\hat{\sigma}_1^y\hat{\sigma}_1^x\hat{\sigma}_1^y - \hat{\sigma}_1^y\hat{n}_1\hat{\sigma}_1^y\,\hat{\sigma}_2^x - \sum_{j = 2}^{L-1} \hat{n}_j\hat{\sigma}_{j+1}^x - \hat{n}_L\\
		= &+\hat{\sigma}_1^x - (\mathbb{I}-\hat{n}_1)\hat{\sigma}_2^x - \sum_{j = 2}^{L-1} \hat{n}_j\hat{\sigma}_{j+1}^x - \hat{n}_L\\
		= & \hat{H}_{-\infty} + 2\hat{\sigma}_1^x - \hat{\sigma}_1^z\hat{\sigma}_2^x
	\end{split}
\end{align}
Thus, the required local perturbation to our Hamiltonian ($\hat{H}_{-\infty} \to \hat{H}_{-\infty} + \hat{\mathcal{O}}$) to create the boundary excitation is
\begin{align}
	\hat{\mathcal{O}} = 2\hat{\sigma}_1^x - \hat{\sigma}_1^z\hat{\sigma}_2^x,
\end{align}

using $\hat{\sigma}^y\hat{n}\hat{\sigma}^y = \mathbb{I}-\hat{n}$ and $2\hat{n}-\mathbb{I} = -\hat{\sigma}^z$. Above perturbation changes the ground state of $\hat{H}_{-\infty}$ to $\ket{\Phi} \equiv -i\hat{\sigma}_1^y \gs$. We perform an $m$-step Lanczos reduction from the initial state $\ket{\Phi}$ to get Ritz values $\theta_j$ with Ritz vectors $\ket{\phi_j}$, where $j = 1, 2, ..., m$ and $m \ll 2^L$. Then, the spectral function can be approximated as
\begin{align}
	\mathcal{A}(\omega) \approx \frac{1}{\pi} \sum_{j = 1}^{m} |\phi_j(1)|^2 \frac{\eta}{(\omega - \omega_j)^2 + \eta^2}
\end{align}
where $\phi_j(1)$ is the first component of the Ritz vector $\ket{\phi_j}$, $\eta$ is a small broadening parameter and $\omega_j \equiv \theta_j - E_\mathrm{GS}$ is the energy gap of the $j$th Ritz value from the ground state energy. We ensure that $\eta$ has the same order of magnitude as the mean level spacing of the Ritz values. The spectral function is normalized such that $\int_{-\infty}^{\infty} \mathcal{A}(\omega) d\omega = 1$. Thus, $\mathcal{A}(\omega)$ can be understood as the local density of states of the state $\ket{\Phi}$.

%================================
%   Finite s
\begin{figure}[t!]
	\centering
	\includegraphics[width=\linewidth]{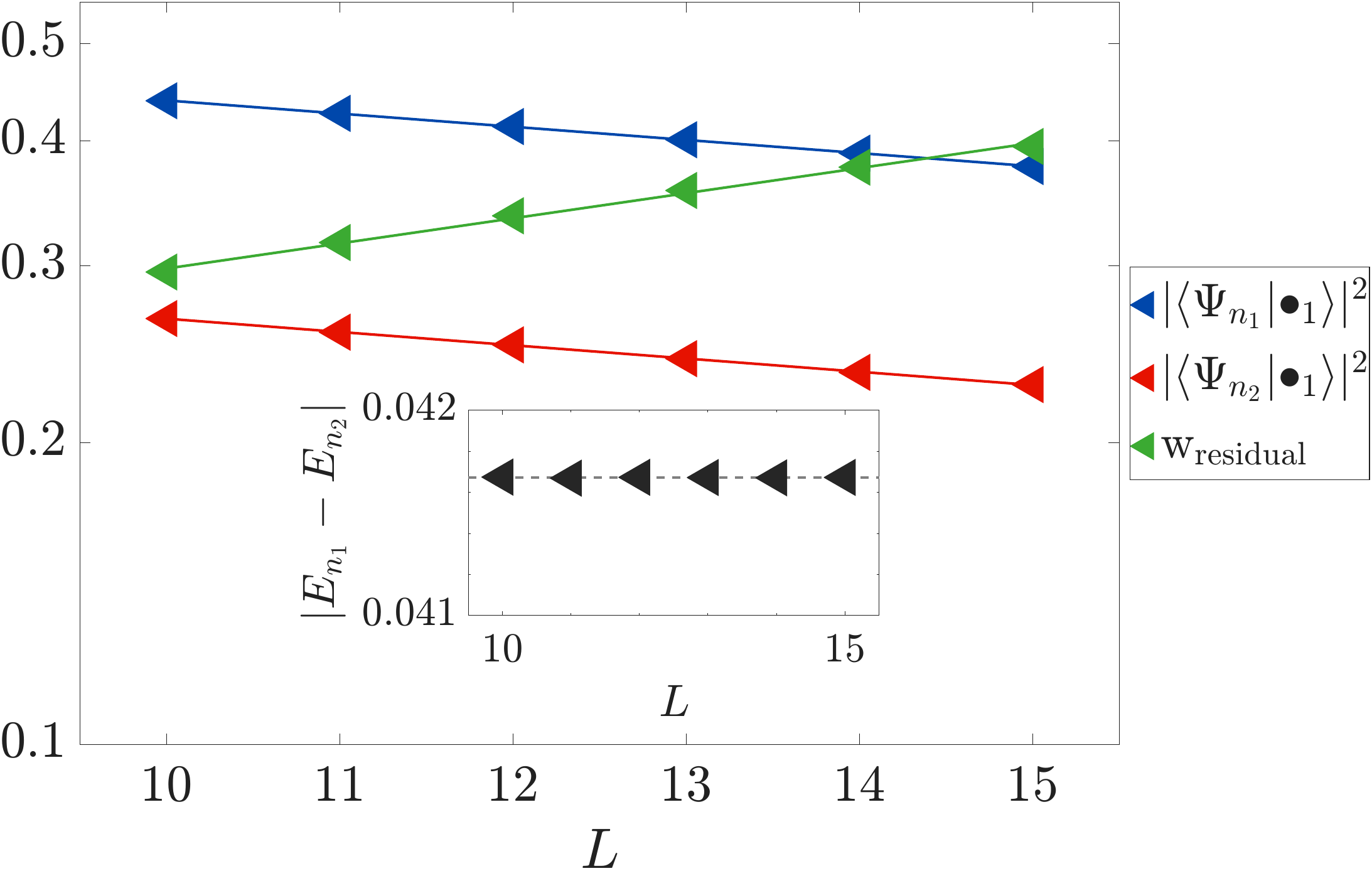}
	\caption{ Two dominant overlaps of $\ket{\bullet_1}$ with eigenstates $\ket{\Psi_{n_{1, 2}}}$ and the residual spectral weight as a function of the system size $L$ on a logarithmic scale for $s = -0.2$. Solid lines are linear fits of the logarithms: both overlaps decrease exponentially, by $3.0\%$ per added site ($\ln|\mean{\Psi_{n_{1}}|\bullet_1}|^2 = -0.521 - 0.0303\,L$ and $\ln|\mean{\Psi_{n_{2}}|\bullet_1}|^2 = -1.020 - 0.0305\,L$), while the residual weight grows by $\approx 6\%$ per site over this range. Inset shows that the energy gap between $\ket{\Psi_{n_{1, 2}}}$ is independent of system size, $|E_{n_1} - E_{n_2}| = 0.04167 \pm 0.00001$ (dashed line: average over $L$).
	}
	\label{fig:esl_olap}
\end{figure}
%================================

In Fig.~\ref{fig:East_L_spectral}(a), we show the spectral function for $L = 20$ and different values of the broadening parameter, $\eta$, chosen between $0.5$ and $2.25$ times the spacing of the Ritz values around the peak. It exhibits a sharp peak at an energy $\omega_0$ whose position is independent of system size, which indicates the presence of a long-lived quasiparticle excitation. To quantify the lifetime of this excitation, we fit the spectral function within $\pm 2\eta$ of the peak to a Lorentzian form ($\mathcal{A}(\omega) \approx \frac{Z}{\pi} \frac{\Gamma}{(\omega - \omega_0)^2 + \Gamma^2}$) and extract the inverse lifetime $\Gamma$, energy $\omega_0$, and weight $Z$ of the quasiparticle peak as a function of $\eta$ for $L = 12, 14, \dots, 20$, as shown in Fig.~\ref{fig:East_L_spectral}(b). The fitted width equals the broadening, $\Gamma = \eta$, for all $\eta$ and $L$, so the intrinsic width extrapolates to zero as $\eta\to 0$: within our resolution the peak is a single eigenstate, $\esl$. The energy $\omega_0 = E_{\mathrm{ES}_1}-E_\mathrm{GS}$ and the weight $Z \approx 0.84$ are independent of $\eta$ and $L$. The extrapolations to $\eta \to 0$ and then $L\to\infty$ (linear in $L^{-1}$) are given in Table~\ref{tab:spectral_fn}. Since $\eta$ cannot be taken below the spacing of the Ritz values, this extrapolation should be regarded as an estimate.
%================================
\begin{table}[t]
	\begin{tabular}{|c|c|c|}
		\hline
		$\Gamma$             & $\omega_0$          & $\mathcal{Z}$        \\ \hline
		$0.0000 \pm 0.0003$ & $2.2598 \pm 0.0002$ & $0.842 \pm 0.003$ \\ \hline
	\end{tabular}
	\caption{Various parameters of the spectral function in the limit $\eta\to 0$ and $L\to\infty$ (from $L = 12, 14, \dots, 20$).}
	\label{tab:spectral_fn}
\end{table}
%================================

Fourier transform of the spectral function gives the return probability of the boundary excitation, $\mathcal{R}(t) = |\langle\Phi|\Phi(t) \rangle|^2 = |\int_{-\infty}^{\infty} \mathcal{A}(\omega) e^{-i\omega t} d\omega|^2$ where $\int_{-\infty}^{\infty} \mathcal{A}(\omega) e^{-i\omega t} d\omega = \mathcal{Z} e^{-i\omega_0 t} e^{-\Gamma t}$. Thus, the excited state has a decay rate $\Gamma$, which here vanishes within our resolution ($\Gamma \ll \omega_0$): the weight $\mathcal{Z}$ of $\ket{\Phi}$ sits on the single eigenstate $\esl$, so $\mathcal{R}(t)$ does not decay below $\mathcal{Z}^2$ on accessible time scales. As the energy bandwidth scales as $\mathcal{O}(\sqrt{L})$, the spectral function is consistent with the presence of a long-lived quasiparticle excitation at the left boundary of the lattice.

Now, we consider the overlap of $\ket{\bullet_1}$ with the eigenstates $\ket{\Psi_n}$ of the East model for finite values of $s$. We find that $\ket{\bullet_1}$ has dominant overlap with two eigenstates denoted by $\ket{\Psi_{n_{1}}}$ and $\ket{\Psi_{n_{2}}}$ with energies $E_{n_1}$ and $E_{n_2}$, respectively. In Fig.~\ref{fig:esl_olap}, we show the dominant overlaps as a function of $L$ for $s = -0.2$. Both overlaps decrease slowly with $L$, by $\approx 3\%$ per site, like the overlap of $\ket{\Theta}$ with the ground state. Let us define the residual weight as $1 - |\langle \Psi_{n_1} | \bullet_1 \rangle|^2 - |\langle \Psi_{n_2} | \bullet_1 \rangle|^2$ which quantifies the contribution of the rest of the eigenstates. In the inset of Fig.~\ref{fig:esl_olap}, we show that the energy gap between the eigenstates $\ket{\Psi_{n_{1}}}$ and $\ket{\Psi_{n_{2}}}$ is approximately independent of system size. We find that the energy gap decreases upon increasing $s$ and vanishes for $s\to 0$.

%===========================================================
%	BIBLIOGRAPHY
\bibliographystyle{apsrev4-2}
\bibliography{ref_Kinetic_constraints}
%===========================================================

\end{document}